\documentclass[oldversion]{aa}
\usepackage{txfonts}
\usepackage{graphicx}
\usepackage{natbib}
\usepackage{amssymb}
\bibpunct{(}{)}{;}{a}{}{,}
\begin{document}
   \title{Mass-loss properties of S-stars on the AGB}
 

   \author{S. Ramstedt\inst{1} \and F. L. Sch\"oier\inst{1} \and H. Olofsson\inst{1,2} \and A.~A. Lundgren\inst{3}}

   \offprints{S. Ramstedt \\ \email{sofia@astro.su.se}}

   \institute{Stockholm Observatory, AlbaNova University Center, SE-106 91 Stockholm, Sweden \and
                   Onsala Space Observatory, SE-439 92 Onsala, Sweden \and European Southern Observatory,
                    Casilla 19001, Santiago 19, Chile }    
             
   \date{Received; accepted}

   \abstract{We have used a detailed non-LTE radiative transfer code to model new APEX CO($J$\,$=$\,3\,$\rightarrow$\,2) data, and existing CO radio line data, on a sample of 40 AGB S-stars. The derived mass-loss-rate distribution has a median value of  2\,$\times$\,10$^{-7}$\,M$_{\odot}$\,yr$^{-1}$, and resembles values obtained for similar samples of M-stars and carbon stars.  Possibly, there is a scarcity of high-mass-loss-rate ($\ge$\,10$^{-5}$\,M$_{\odot}$\,yr$^{-1}$) S-stars. The distribution of envelope gas expansion velocities is similar to that of the M-stars, the median is 7.5 km\,s$^{-1}$, while the carbon stars, in general, have higher gas expansion velocities. The mass-loss rate correlates well with the gas expansion velocity, in accordance with results for M-stars and carbon stars.
   
   \keywords{Stars: AGB and post-AGB -- Stars: circumstellar matter -- Stars: late-type -- Stars: mass-loss}
   }
   \maketitle
%

\section{Introduction}
Stars of spectral class S are identified through their strong bands of primarily ZrO. This indicates enhancements of s-process elements. It is now relatively well-established that these enhancements can be produced in two ways: through dredge-up of nuclear-processed material during the asymptotic giant branch (AGB) evolution (intrinsic S-star), or through mass transfer in a binary system (extrinsic S-star). 
The strong bands of ZrO in intrinsic S-stars, as opposed to the strong bands of TiO in M-stars, have been attributed to a stellar atmosphere chemistry in a relatively low-temperature ($\la$\,3000\,K) gas 
enriched in s-process elements and with C/O\,$=$\,1 (within about 5\%) \citep{scalross76,smitlamb90}. This then lead to the belief that the AGB S-stars are transition objects between AGB M-stars and carbon stars.

Extensive mass loss is a key process during the evolution of AGB stars. M-stars and carbon stars lose substantial amounts of matter during this phase, and their mass-loss characteristics are relatively well-established \citep[see][]{Olofsson03}. The putative transition objects, the S-stars, have attracted less attention in this respect. Studies of CO radio line and dust continuum emission suggest that they behave in roughly the same way as both their predecessors and descendants in terms of mass-loss rate, circumstellar kinematics, dust composition, and dust-to-gas ratio \citep{jura88,bieglatt94,sahaliec95,groedejo98,Jorissen98}.

In this {\em Letter} new APEX\footnote{This publication is based on data acquired with the Atacama Pathfinder Experiment (APEX). APEX is a collaboration between the Max-Planck-Institut f\"{u}r Radioastronomie, the European Southern Observatory, and the Onsala Space Observatory.} CO($J$\,$=$\,3\,$\rightarrow$\,2) data, in combination with existing CO radio line data and a radiative transfer code, were used to derive the circumstellar properties of 40 S-stars on the AGB.

   \begin{figure*}
   \centering{   
   \includegraphics[width=17cm]{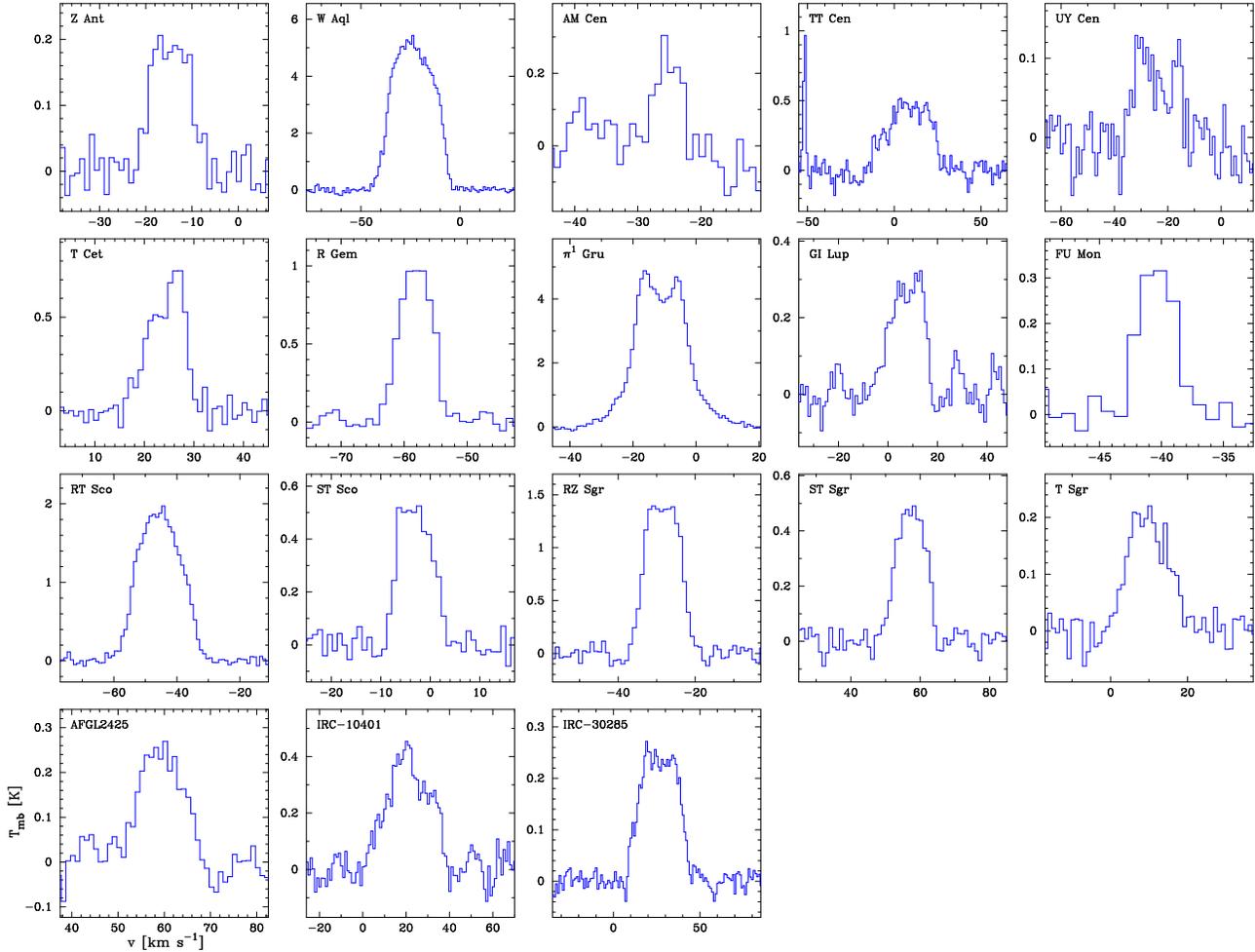}
   \caption{New observations of CO($J$\,$=$\,3$\rightarrow$\,2) line emission using the APEX telescope. The velocity resolution is 1.0\,km\,s$^{-1}$.}
   \label{obs_apex}}
   \end{figure*}

\section{Observations}
\label{sect_obs}
The sample was selected from the list of 65 objects with S-star characteristics in the {\em Two Micron Sky Survey} by \citet{wingyork77} [see also \citet{jura88}]. Observations of CO $J$\,$=$\,3\,$\rightarrow$\,2 line emission at 345.80\,GHz towards a sample of 18 S-stars were performed from August to October 2005 using the APEX 12\,m telescope equipped with a double-sideband SIS receiver and a FFTS spectrometer. The system temperature (above atmosphere)
varied between 130 and 335\,K. The effective bandwidth at the observations were 1\,GHz with a spectral resolution of 61\,kHz (0.05\,km\,s$^{-1}$), except for FU~Mon, R~Gem, and Z~Ant observed during the second run with a spectral resolution of 122\,kHz (0.11\,km\,s$^{-1}$). The data was smoothed to 1\,km\,s$^{-1}$ resolution and reduced by removing a low-order ($\leq$\,3) polynomial baseline in the range of about $\pm$\,100 km/s from the stellar velocity. The resulting spectra are presented in Figure.~\ref{obs_apex}, and velocity-integrated intensities are reported in Table~\ref{sample}. 

The observations were carried out using a position-switching mode, with the reference position located +2$\arcmin$ in azimuth. The raw spectra are stored in the $T_{\mathrm A}^{\star}$  scale and converted to main-beam brightness temperature using $T_{\mathrm{mb}}$\,=\,$T_{\mathrm A}^{\star}/\eta_{\mathrm{mb}}$, where $T_{\mathrm A}^{\star}$ is the antenna temperature corrected for atmospheric attenuation using the chopper-wheel method, and $\eta_{\mathrm{mb}}$ the main-beam
efficiency, estimated to be 0.7 at 345\,GHz. The uncertainty of the absolute intensity scale is estimated to be about $\pm$\,20\%. 
Regular pointing checks were made on strong CO sources and typically found to be consistent within $\approx$\,3$\arcsec$. The FWHM of the main beam, $\theta_{\mathrm{mb}}$, is 18$\arcsec$ at 345\,GHz. 

In addition to the new APEX data presented here, we have also used new CO($J$\,$=$\,1\,$\rightarrow$\,0) data observed at the Onsala 20\,m telescope ( January 2006), JCMT\footnote{The James Clerk  Maxwell Telescope, is operated by the Joint Astronomy Centre in Hilo,  Hawaii on behalf of the parent organizations PPARC in the United  Kingdom, the National Research Council of Canada and The Netherlands  Organization for Scientific Research.} public archive CO data, and CO line intensities reported by \citet{bieglatt94}, \citet{sahaliec95}, \citet{Stanek95}, \citet{groedejo98}, and \citet{Jorissen98}. In total, 42 S-stars have been detected in CO line emission. 

\section{Radiative transfer analysis}
\label{sect_model}
In our analysis the circumstellar envelopes (CSEs) are assumed to be spherically symmetric, to be produced by a constant mass-loss rate ($\dot{M}$), and to expand at a constant velocity ($v_{\mathrm e}$). The spectra of $\pi^{1}$~Gru and RS~Cnc indicate that deviations from the standard CSE model are present and no modelling was attempted for these two sources
\citep[see][]{saha92,knapetal99a}. 
\subsection{SED modelling}
The SEDs were constructed from IRAS-fluxes and 2MASS-magnitudes. The continuum emission was modelled using the dust radiative transfer code DUSTY \citep{Ivezic97}. 
Amorphous carbon or amorphous silicate dust was selected based on, in most cases, the spectral class of the IRAS low-resolution spectra as defined by \citet{Volk89}. In the cases where the spectral class was not available, the selection was based on colour-colour diagrams as described by \citet{Jorissen98}. The adopted dust type is listed in Table~\ref{sample}, where C and O denotes carbon and silicate dust, respectively. 
The amount of dust present in the wind can be constrained from the spectral shape of the SED. Details on the modelling procedure can be found in \citet{Schoeier02b} and \citet{Ramstedt06b}. 

The SED results were used in combination with the luminosities to estimate the
distances to the sample stars. Luminosities of the Miras were estimated using the period-luminosity relation of \citet{Whitelock94}. For the semi-regular and irregular variables, a luminosity of 4000\,$\mathrm{L_{\odot}}$ was assumed in accordance with \citet{olofetal02}. Corrections for interstellar extinction were applied. The derived distances and adopted luminosities are listed in Table~\ref{sample}. Non-negative Hipparcos parallaxes with an accuracy better than 50\,\% are available for seven of the sample stars. They agree well within the error margins, when compared to the distances derived from the SED modelling (except for RZ Peg).

The results from the SED modelling were also used to determine the dust radiation field which is important for the excitation of CO.

\begin{table}
\caption{APEX observations and model summary}
\label{sample}
\resizebox{\hsize}{!}{ 
$
\begin{array}{p{0.162\linewidth}cccccccccc}
\hline
\noalign{\smallskip}

& & & &

\multicolumn{6}{c}{\mathrm{CO\ modelling}}\\
\noalign{\smallskip}
\cline{5-10}
\noalign{\smallskip}
\multicolumn{1}{c}{{\mathrm{Source}}} &
\multicolumn{1}{c}{D}& 
\multicolumn{1}{c}{L_{\star}}&
\multicolumn{1}{c}{I(\mathrm{3-2})} &

\multicolumn{1}{c}{\dot{M}}&
\multicolumn{1}{c}{v_{\mathrm{e}}} &
\multicolumn{1}{c}{h} &
\multicolumn{1}{c}{\chi^2_{\mathrm{red}}} &
\multicolumn{1}{c}{N} & 
\multicolumn{1}{c}{{\mathrm{dust}}} \\
&
\multicolumn{1}{c}{[\mathrm{pc}]}  &
\multicolumn{1}{c}{[\mathrm{L_{\odot}}]} &  
\multicolumn{1}{c}{[\mathrm{Kkms^{-1}}]} &  
\multicolumn{1}{c}{[\mathrm{M_{\odot}\,yr^{-1}}]} &
\multicolumn{1}{c}{[\mathrm{kms^{-1}}]} & 
& & &
\multicolumn{1}{c}{{\mathrm{type}}} \\
\noalign{\smallskip}
\hline
\noalign{\smallskip}
\object{R And}\,$^{b}$             & 300 & 6000  &\cdots & 6.0\times 10^{-7} & 8.0 & 0.2$:$ & 0.1 & 2 & $O$    \\
\object{W And}\,$^{a}$            & 280 & 5800  &\cdots & 1.2\times 10^{-7} & 5.5 & 0.2$:$ & 1.0  & 3 & $O$ \\
\object{Z Ant}               & 470 & 4000  & 2.0 & 9.0\times 10^{-8} & 6.0 & 0.2$:$ & \cdots & 1 & $O$  \\
\object{VX Aql}           & 790 & 4000  & \cdots  & 5.0\times 10^{-8} & 7.0$:$ & 0.2$:$ & \cdots & 1 & $C$ \\
\object{W Aql\,$^{c}$ }              & 230 & 6800  & 136  & 2.5\times 10^{-6} & 17.5 & 0.7 &  2.5 & 4 & $O$ \\
\object{AA Cam}              & 800 & 4000  & \cdots & 1.8\times 10^{-7} & 16.9$:$ & 0.2$:$ & \cdots & 1 & $C$ \\
\object{T Cam}              & 540 & 5600  & \cdots & 1.0\times 10^{-7} & 4.8$:$ & 0.2$:$ & \cdots & 1 &  $C$\\
\object{S Cas}\,$^{a,b}$             & 440 & 8000  &\cdots & 2.2\times 10^{-6} & 19.0 & 0.5$:$ & 0.9 & 2 & $O$ \\
\object{V365 Cas}              & 625 & 4000  & \cdots & 3.0\times 10^{-8} & 6.2$:$ & 0.2$:$ & \cdots & 1 &$C$ \\
\object{WY Cas}              & 600 & 6700  & \cdots & 6.0\times 10^{-7} & 12.5$:$ & 0.5$:$ & \cdots & 1 & $O$ \\
\object{AM Cen}              & 750 & 4000  & 1.1 & 1.5\times 10^{-7} & 3.3 & 0.2$:$ & 1.9 & 2 & $C$ \\
\object{TT Cen}              & 880 & 6500  & 13.9 & 2.5\times 10^{-6} & 20.0 & 0.5$:$ & 7.8 & 2 & $O$ \\
\object{UY Cen}           & 590 & 4000  & 1.6  & 1.2\times 10^{-7} & 12.0 & 0.2$:$ & \cdots & 1 & $C$ \\
\object{V386 Cep}              & 470 & 4000  & \cdots & 1.8\times 10^{-7} & 18.6$:$ & 0.2$:$ & \cdots & 1 & $O$ \\
\object{T Cet}\,$^{a,b,c,d}$               & 240 & 4000 & 3.7 & 5.0\times 10^{-8} & 5.5 & 0.2 & 0.8 & 4 &  $C$ \\
\object{AA Cyg}              & 480 & 4000  & \cdots & 4.0\times 10^{-7} & 6.1$:$ & 0.2$:$ & \cdots & 1 & $C$  \\
\object{R Cyg}\,$^{a}$         & 440 & 6100  & \cdots & 5.0\times 10^{-7} & 9.5 & 0.5$:$ & 1.8 & 3 & $O$ \\
\object{$\chi$ Cyg}\,$^{a,b,c}$    & 110 & 5900  &\cdots & 5.0\times 10^{-7}  & 8.5 & 0.1 & 2.0 & 3 &  $O$ \\
\object{TV Dra}\,$^{a}$            & 390 & 4000  & \cdots  & 7.0\times 10^{-8} & 4.0 & 0.2$:$ & \cdots & 1 &  $O$ \\
\object{DY Gem}         & 680 & 4000 &\cdots & 1.3\times 10^{-7} & 7.0 & 0.2$:$ & \cdots  & 1&  $O$ \\
\object{R Gem}\,$^{a}$    & 710 & 5500  & 6.4 & 4.0\times 10^{-7} & 4.0 & 0.5$:$ & 1.0 & 2 & $C$ \\
\object{ST Her}\,$^{a}$            & 300 & 4000  &\cdots & 1.3\times 10^{-7} & 8.5 & 0.2$:$ & 1.1 & 3 & $O$ \\
\object{RX Lac}              & 310 & 4000  & \cdots & 6.5\times 10^{-8} & 5.5$:$ & 0.2$:$ & \cdots & 1 &  $C$ \\
\object{GI Lup}              & 690 & 5000  & 4.6 & 1.0\times 10^{-6} & 10.0 & 0.2$:$ & \cdots & 1 &  $C$ \\
\object{R Lyn}\,$^{a}$             & 850 & 5600  &\cdots & 3.0\times 10^{-7} & 7.0 & 0.2$:$ & 0.2  & 2 &  $C$ \\
\object{Y Lyn}\,$^{a}$         & 260 & 4000  &\cdots  & 1.7\times 10^{-7} & 7.0 & 0.2$:$ & \cdots & 1 & $O$ \\
\object{S Lyr}              & 1210 & 6300  & \cdots & 1.0\times 10^{-6} & 13.0$:$ & 0.5$:$ & \cdots & 1 &  $O$ \\
\object{FU Mon}          & 780 & 4000  & 1.3  & 1.2\times 10^{-7} & 2.0 & 0.2$:$ & \cdots & 1 &  $C$ \\
\object{RZ Peg}	& 970 & 4000 & \cdots & 3.4\times 10^{-7} & 12.6$:$ & 0.2$:$ & \cdots & 1 & $C$ \\
\object{RT Sco}           & 270 & 6400  & 32.6  & 4.5\times 10^{-7} & 11.0 & 0.5$:$ & 3.2 & 2 & $O$ \\
\object{ST Sco}           & 380 & 4000  & 4.5  & 1.3\times 10^{-7} & 5.5 & 0.2$:$ & 1.2 & 3 &  $O$ \\
\object{RZ Sgr}              & 730 & 4000  & 15.8 & 1.5\times 10^{-5} & 9.0 & 0.2$:$ & 22.3 & 2 & $C$ \\
\object{ST Sgr}              & 540 & 5800  & 4.9 & 2.7\times 10^{-7} & 7.0 & 0.2$:$ & 3.0 & 2 & $O$ \\
\object{T Sgr}              & 590 & 5800  & 2.5 & 2.0\times 10^{-7} & 10.0 & 0.2$:$ & \cdots & 1 & $C$ \\
\object{EP Vul}              & 510 & 4000  & \cdots & 1.5\times 10^{-7} & 4.4$:$ & 0.2$:$ & 5.5 & 2 &  $C$ \\
\object{DK Vul}              & 750 & 4000  & \cdots & 6.0\times 10^{-7} & 4.0$:$ & 0.2$:$ & \cdots & 1 &  $C$ \\
\object{AFGL2425}              & 610 & 4000  & 2.8 & 2.0\times 10^{-7} & 6.5 & 0.2$:$ & 0.9 & 2 &  $O$\\
\object{CSS2 41}	& 880 & 4000 & \cdots &  6.0\times 10^{-7} &  20.0$:$ &  0.2$:$ & \cdots & 1 & $O$ \\
\object{IRC-30285}              & 590 & 4000  & 6.5 & 6.0\times 10^{-7} & 17.0 & 0.2$:$ & 6.0 & 2 &  $O$ \\
\object{IRC-10401}              & 430 & 4000  & 9.5 & 3.5\times 10^{-7} & 17.0 & 0.2$:$ & 12.8 & 2 & $O$\\

\noalign{\smallskip}
\hline
\end{array}
$
}
Superscripts mark new or archive data: $^{a}$\,OSO $J$\,=\,1\,$\rightarrow$\,0,\\ $^{b}$\,JCMT $J$\,=\,3\,$\rightarrow$\,2 , $^{c}$\,JCMT $J$\,=\,4\,$\rightarrow$\,3, $^{d}$\,JCMT $J$\,=\,6\,$\rightarrow$\,5.\\ The forth
column gives the APEX $J$\,=\,3\,$\rightarrow$\,2 results.
\end{table}
  \begin{figure*}
    \centering{   
    \includegraphics[width=17.5cm]{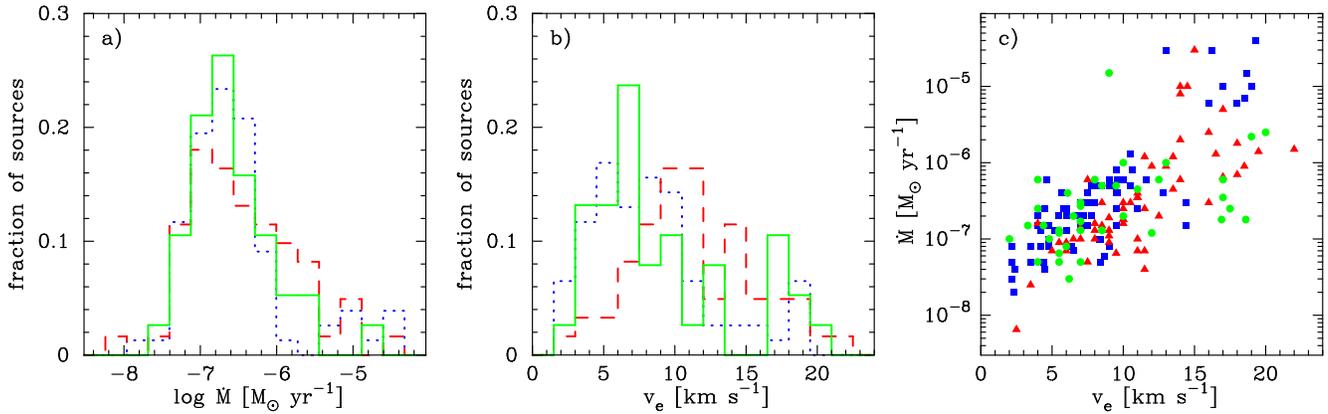}
    \caption{{\bf a)} Mass-loss-rate distributions for the S-star (solid
    line; 40 stars), M-star (dotted line; 77 stars), and carbon star (dashed line; 61 stars) samples.
    {\bf b)} Envelope gas expansion velocity distributions for the S-star (solid
    line), M-star (dotted line), and carbon star (dashed line) samples.
    {\bf c)} Derived mass-loss rates plotted against the gas expansion velocities
    for the S-star (dots), M-star (squares), and carbon star (triangles) samples.}
          \label{histM}}
   \end{figure*}

\subsection{Radiative transfer model}
In order to determine the molecular excitation in the CSE, a detailed non-LTE radiative transfer code, based on the Monte Carlo method,
was used.  This code is described in detail in \citet{Schoeier01}. It has been
benchmarked, to high accuracy, against a wide variety of molecular-line radiative
transfer codes in \citet{Zadelhoff02}.  

The CO molecules are partly excited through collisions with H$_2$. The collisional rate coefficients calculated by \citet{Flower01a} were adopted and extended to include more energy levels as well as extrapolated in temperature \citep[see][]{Schoeier05a}. An ortho-to-para ratio of 3 was adopted when weighting together collisional rate coefficients for CO in collisions with ortho- and para-H$_2$.
The excitation analysis includes radiative excitation through the first vibrationally excited ($v$\,=\,1) state  at 4.6\,$\mu$m. Relevant molecular data are summarised in \citet{Schoeier05a} and are made publicly available through the {\em Leiden Atomic and Molecular Database} (LAMDA){\footnote{\tt http://www.strw.leidenuniv.nl/$\sim$moldata}}. A central blackbody was assumed to represent the stellar radiation field, and thermal, circumstellar dust emission was included. These provide the main sources of the infrared photons that excite the $v$\,=\,1 state. The addition of a dust component in the Monte Carlo scheme is straightforward as described in \citet{Schoeier02b}. The energy-balance equation was solved simultaneously as the excitation of CO, allowing for an accurate determination of the kinetic gas temperature of the gas. The photospheric abundance of CO, relative to H$_{2}$, was assumed to be 6\,$\times$\,10$^{-4}$. A micro-turbulent velocity of 1 km s$^{-1}$ was assumed. 

Two parameters are adjustable when fitting the model to the observed line intensities: the mass-loss rate ${\dot{M}}$ and the $h$-parameter. The latter is a product of different dust parameters \citep{Schoeier02b}. Collisional heating through gas colliding with dust grains is the dominant heating term in the energy-balance equation and it affects the model line-intensity ratios. We were able to constrain $h$ for five of our stars. Its dependence on stellar luminosity is consistent
with our findings for M-stars \citep{olofetal02} and
carbon stars \citep{Schoeier01}. Therefore, a value was assumed for the remaining stars depending on the stellar luminosity, $h$=0.2 for $L$\,$<$\,5000\,L$_{\odot}$ and $h$=0.5 above 5000\,L$_{\odot}$ (indicated by a colon in Table~\ref{sample}).

The best-fit model was found by minimizing the difference between the observed and modelled integrated line intensities ($N$, in Table~\ref{sample}, is the number of observed lines used). Reduced $\chi^{2}$-values were calculated and are presented in Table~\ref{sample}, when a sufficient number of lines were available. The uncertainties in the derived mass-loss rates were estimated to be about $\pm$\,50\% within the adopted circumstellar model, excluding distance uncertainties, in accordance with \citet{Schoeier01} and \citet{olofetal02}.

A particular problem occurred with the CO($J$\,=\,2\,$\rightarrow$\,1) data. We found a large scatter in the reported intensities for an individual object, even when observed with the same telescope. Furthermore, these data are, in general, not compatible with the model results, i.e., the 3\,$\rightarrow$\,2/1\,$\rightarrow$\,0 intensity ratios are compatible with the model results, while the 2\,$\rightarrow$\,1/1\,$\rightarrow$\,0 or 3\,$\rightarrow$\,2/2\,$\rightarrow$\,1 intensity ratios are, in general, not (note, the model line-intensity ratios only weakly depend on the assumed parameters).
Consequently, we decided to omit the 2\,$\rightarrow$\,1 data from our analysis, except for the few cases
where only a 2\,$\rightarrow$\,1 line exists.
 
\section{Results and discussion}
\label{sect_discussion}

We compare here our results for the S-star sample, which were obtained through detailed radiative transfer modelling, with those obtained in the same way for a sample of M-stars \citep{olofetal02} and a sample of carbon stars \citep{Schoeier01}. The samples are flux-limited but not complete, except for the carbon-star sample, which contains a subsample that is most likely complete out to a distance of 500\,pc \citep{Schoeier01}. The incompleteness of the samples and the small number statistics limit the generality of our conclusions.

Figure~\ref{histM}a shows the mass-loss-rate distribution of the S-star sample in comparison with those of the M-stars and the carbon stars. The S-star mass-loss-rate distribution appears very similar to that of the M-stars and the carbon stars. The median value is 2\,$\times$\,10$^{-7}$\,M$_{\odot}$\,yr$^{-1}$ for the M- and S-star samples, and 3\,$\times$\,10$^{-7}$\,M$_{\odot}$\,yr$^{-1}$ for the carbon-star sample. There is possibly fewer S-stars with a high mass-loss rate ($\ge$\,10$^{-5}$\,M$_{\odot}$\,yr$^{-1}$).
Figure~\ref{histM}b shows the expansion velocity distribution of the S-star sample in comparison with those of the M-stars and carbon star samples. There is an indication that the carbon stars have higher expansion velocities, while those of the S- and M-stars appear similar. The median expansion velocity is 7.5\,km s$^{-1}$ for the S-star and M-star samples, and 11\,km s$^{-1}$ for the carbon star sample. 
Figure~\ref{histM}c shows that the relations between mass-loss rate and
envelope expansion velocity are very similar for the three samples, pointing towards
the same mass-loss mechanism.
 
This is the first step in a project aimed at obtaining an extensive data base of circumstellar molecular lines towards a sample of S-stars. The CO data will constrain the circumstellar
model, and accurate molecular abundances will be derived. These will be compared with results of chemical modelling, as well as with the results for the M-star and carbon star samples.  
  
\begin{acknowledgements}
SR, FLS, and HO acknowledge financial support from the Swedish Research Council.
\end{acknowledgements}

\bibliographystyle{aa}

\end{document}